\documentclass[aps,prb,twocolumn,showpacs]{revtex4}
\usepackage{amsmath}

\usepackage{graphicx}

\begin{document}

\title{Band-Contact Lines in Electron Energy Spectrum of Graphite}

\author{G.~P.~Mikitik}

\author{Yu.~V.~Sharlai}

\affiliation{B.~Verkin Institute for Low Temperature Physics \&
  Engineering, Ukrainian Academy of Sciences,
   Kharkov 61103, Ukraine}

\date{\today}

\begin{abstract}
We discuss the known experimental data on the phase of the de Haas
-van Alphen  oscillations in graphite. These data can be
understood if one takes into account that four band-contact lines
exist near the HKH edge of the Brillouin zone of graphite.
\end{abstract}


\maketitle

\section{Introduction}

At present, graphite and its electronic properties attract
considerable attention due to the discovery of novel carbon-based
materials such as fullerenes and nanotubes constructed from
wrapped graphite sheets. \cite{1} Besides, thin films of graphite
give promise of device applications. \cite{N1} The attention to
graphite is also caused by specific features of its electron
energy spectrum  which result in interesting physical effects.
\cite{K1,N2} The electronic spectrum of graphite is described by
the Slonzewski-Weiss-McClure (SWM) model, \cite{SW,Mc} and values
of the main parameters of this model were found sufficiently
accurately from an analysis of various experimental data; see,
e.g., the review of Brandt {\it et al}. \cite{Br} and references
therein. The Fermi surface of graphite consists of elongated
pockets enclosing the edge HKH of its Brillouin zone ( see figures
below). These pockets are formed by the two majority groups of
electrons (e) and holes (h) which are located near the points K
and H of the Brillouin zone, respectively. There is also at least
one minority (m) low-concentration group of charge carriers in
graphite, and this group seems to be located near the point H.
However, it is necessary to emphasize that in spite of the
considerable attention attracted to graphite an unresolved problem
concerned with its spectrum still exists.

It is well known \cite{SW,Mc,Br} that in the edge HKH of the
Brillouin zone of graphite, two electron energy bands are
degenerate, and in a small vicinity of the edge these bands split
linearly in a deviation of the wave vector ${\bf k}$ from the
edge. In other words, the edge is the band-contact line. But, it
was shown in our paper \cite{prl} that if in the ${\bf k}$-space a
closed semiclassical orbit of a charge carrier surrounds a contact
line of its band with some other band (and lifting of the
degeneracy is linear in ${\bf k}$), the wave function of this
carrier after its turn over the orbit acquires the addition phase
$\Phi_B=\pm \pi$ as compared to the case without the band-contact
line. This $\Phi_B$ is the so-called Berry phase, \cite{Ber} and
it modifies the constant $\gamma$ in the well-known semiclassical
quantization rule \cite{Sh} for the energy $\varepsilon$ of a
charge carrier in the magnetic fields $H$:
\begin{equation}\label{1}
S(\varepsilon ,k_H)=\frac{2\pi e H}{\hbar c}(n+\gamma ),
\end{equation}
where ${S}$ is the cross-sectional area of the closed orbit in the
${\bf k}$ space; $k_H$ is the component of ${\bf k}$ perpendicular
to the plane of this orbit; ${n}$ is a large integer (${n>0}$);
${e}$ is the absolute value of the electron charge, and the
constant $\gamma$ is now given by the formula:\cite{prl}
\begin{equation}\label{2}
\gamma = \frac{1}{2} -\frac{\Phi_B}{2\pi}\ .
\end{equation}
When the magnetic field is applied along the HKH axis, orbits of
electrons and holes in the Brillouin zone of graphite surround
this axis. Thus, one might expect to find $\gamma=0$ for these
orbits instead of the usual value $\gamma=1/2$ (the values
$\gamma=0$ and $\gamma=1$ are equivalent).

A value of $\gamma$ can be measured using various oscillation
effects and in particular, with the de Haas - van Alphen effect.
\cite{zhetf,g,prl1} For example, the first harmonic of the de Haas
- van Alphen oscillations of the magnetic susceptibility has the
form, \cite{LK}
\begin{equation}\label{3}
 \chi \cos\left (2\pi\frac{\nu}{H}+\phi \right),
\end{equation}
where $\nu=\hbar c S_{ex}/(2\pi e)$, $S_{ex}$ is some extremal
cross section of the Fermi surface of a metal in $k_H$, a positive
$\chi$ is the amplitude of this first harmonic, and $\phi$ is its
phase which is given by
\begin{equation}\label{4}
 \phi = -2\pi \gamma +\delta
\end{equation}
with $\delta = \pm \pi/4$ for a minimum and maximum cross-section
$S_{ex}$, respectively, and $\delta=0$ in the case of a
two-dimensional Fermi surface. \cite{K} It follows from
Eq.~(\ref{4}) that one has to obtain $\phi_e=\phi_h=-\pi/4$ for
the maximum cross-sections of the electron and hole majorities in
graphite. However, the phases $\phi_e$, $\phi_h$ measured long ago
\cite{S,W} agree with the usual value $\gamma=0.5$; see Table I.

Recently a new method of determining the phase $\phi$ of the de
Haas - van Alphen oscillations was elaborated, \cite{K} and the
authors of that paper found $\gamma=0$ for the cross section of
the hole majority in graphite. However, in this determination they
assumed the Fermi surface of the holes to be two-dimensional
($\delta_h=0$); see Table I. Besides this, they found $\gamma=0.5$
for the maximum cross section of the electron majority, assuming
the three-dimensional Fermi surface for this majority
($\delta_e=-\pi/4$). Although the obtained value $\gamma=0$ for
the holes agrees with the above prediction, the results of
Ref.~\onlinecite{K} give rise to the following new problems:
First, since the band-contact line in graphite penetrates both the
electron and hole extremal cross sections, these cross sections
must have the same $\gamma$. Second, using the values of the
parameters of SWM model, \cite{Br} one might expect that in
graphite the electrons and holes of the extremal cross sections
are both three-dimensional.

In this paper we show that in graphite, apart from the
band-contact line coinciding with the edge HKH, {\it three
additional} band-contact lines exist near this edge. The existence
of these lines leads to the usual value $\gamma=0.5$ for the
maximum cross sections of the electron and hole majority groups in
graphite. In other words, we resolve the above-mention
contradiction between the theoretical value of $\gamma$ and the
data of Refs.~\onlinecite{S,W}. We also discuss the data of
Ref.~\onlinecite{K}.

\begin{table*}
\caption{\label{TI}Frequencies $\nu$, phases $\phi$, and the
appropriate $\gamma$ and $\delta$ of quantum oscillations in
graphite for $S_e$, $S_h$, $S_m$ of Fig.~2}
\begin{ruledtabular}
\begin{tabular}{l|crrc|crrc|cl}
 & \multicolumn{4}{c|}{ WFD \cite{W}$^,$\cite{Br}}
 & \multicolumn{4}{c|}{LK \cite{K} }
 & \multicolumn{2}{c}{LK corrected } \\
\hline
  & $\nu$ (kOe)
      & $(\phi/\pi)$
      & $(\delta/\pi)$ & $\gamma$
  & $\nu$ (kOe) & $(\phi/\pi)$ & $(\delta/\pi)$ & $\gamma$
  & $(\delta/\pi)$
      & $\gamma$
                     \\
\hline
%majority
$e$ &  65$\pm$4 & 0.64$\pm$0.18 & -1/4 & 1/2
  & 46.8 & 3/4 & -1/4 & 1/2
  & -1/4 & 1/2\\
%majority
$h$ &  46$\pm$3 & 0.76$\pm$0.1 & -1/4 & 1/2
  & 64.1 &   1   & 0        & 0
  & -1/4 & 3/8\\
%minority
$m$ & 6$\pm$3 & 0.06 & 0 & 0
  & 3.28 & 0         & 0        & 1/2
  & 0        & 0  \\
\end{tabular}
\end{ruledtabular}
\end{table*}

 \begin{figure}  % 1 %%%%%%%%%%%%%%%%%%%%%%%%%%%%%%%%%%%%%%%%
 \includegraphics[scale=.97]{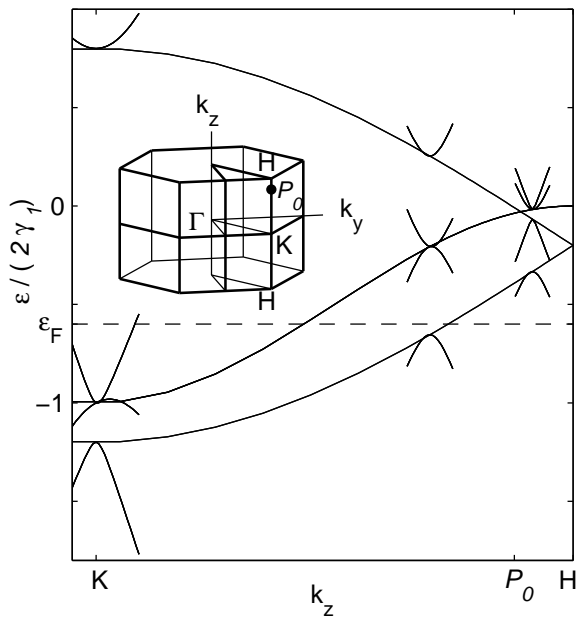}
\caption{\label{fig1} Dependences of the bands $\varepsilon_i$
($i=1-4$) on ${\bf k}$ near the edge HKH in graphite. Shown are
the dependences $\varepsilon_i(k_z)$ at $k_\bot=0$ and the
dependences of $\varepsilon_i$ on $k_\bot=\sqrt{k_x^2+k_y^2}$ at
some characteristic values of $k_z$. The dashed line marks the
position of the Fermi level $\varepsilon_F$. For clarity, in the
construction of the figure the parameter $\gamma_1$ has been used
which is twenty times smaller than that of Table II. For realistic
value of $\gamma_1$ the point $P_0$ is much closer to the point H
than in the figure. The insert shows the Brillouin zone of
graphite and its characteristic points.
 } \end{figure}   %%%%%%%%%%%%%%%%%%%%%%%%%%%%%%%%%%%%%%%%%%

\section{Band-contact lines in graphite}

The SWM model \cite{SW,Mc} describes the wave-vector dependence of
four electron energy bands of graphite $\varepsilon_i({\bf k})$
($i=1-4$) in the vicinity of the vertical edge HKH of its
Brillouin zone, Fig.~1. These bands can be found from the
forth-order secular equation:
\begin{equation}\label{5}
\rm{ det} \left | \hat{H} - \varepsilon \right | =0,
\end{equation}
where the Hamiltonian matrix $\hat{H}$ has the form
\begin{equation}\label{6}
\hat{H}=
\left (
    \begin{array}{cccc}
    E_1 & 0 & H_{13} & H_{13}^{*} \\
    0 & E_2 & H_{23} & -H_{23}^{*} \\
    H_{13}^{*} & H_{23}^{*} & E_3 & H_{33} \\
    H_{13} & -H_{23} & H_{33}^{*} & E_3
    \end{array}
\right ).
\end{equation}
Here the following notations have been used:
\begin{eqnarray}
E_1=\Delta+\gamma_1\Gamma+\frac 12 \gamma_5\Gamma^2, \nonumber \\
E_2=\Delta-\gamma_1\Gamma+\frac 12 \gamma_5\Gamma^2, \nonumber \\
E_3=\frac 12 \gamma_2 \Gamma^2, \label{7} \\
H_{13}=\frac 1{\sqrt{2}}  (
   -\gamma_0+\gamma_4\Gamma
 ) {\rm e}^{i\alpha}\zeta, \nonumber \\
H_{23}=\frac 1{\sqrt{2}} \left (
    \gamma_0+\gamma_4\Gamma,
\right ) {\rm e}^{i\alpha} \zeta, \nonumber \\
H_{33}=\gamma_3\Gamma{\rm e}^{i\alpha}\zeta, \nonumber
\end{eqnarray}
where $\alpha$  is the angle between the direction of the vector
${\bf k}$ and the $\Gamma K$ direction in the Brillouin zone;
$\Gamma=2\cos \xi$; $\xi$ and $\zeta$ are dimensionless wave
vectors in the direction of the $z$-axis and in the basal plane,
respectively: $\xi=(\pi/ 2)( k_z /|KH|)$, $\zeta=( 2\pi
/\sqrt{3})(k_\bot/|\Gamma K|)$; $k_\bot=\sqrt{k_x^2+k_y^2}$; ${\bf
k}$ is measured from the point K. The parameter $\gamma_0$ which
describes the interaction between neighbor atoms in a graphite
layer is sufficiently large as compared to the other parameters
$\gamma_i$, $\Delta$ which describe interactions between atoms in
different graphite layers; see Table II. It is known \cite{Br}
that the band structure near the point $H$ is highly sensitive to
values of the small parameters  $\gamma_2$ and $\Delta$. While the
value of $\gamma_2$ is known sufficiently well, the value of
$\Delta$ was found less reliably. \cite{Br} Although our main
conclusions remain unchanged for any reasonable $\Delta$, for
definiteness, in subsequent analysis we shall use the set of the
parameters \cite{Br} based on the data of
Refs.~\onlinecite{S,T,M,D,Me1,Me2} and presented in Table II.

 \begin{figure}  % 2 %%%%%%%%%%%%%%%%%%%%%%%%%%%%%%%%%%%%%%%%
 \includegraphics[scale=.97]{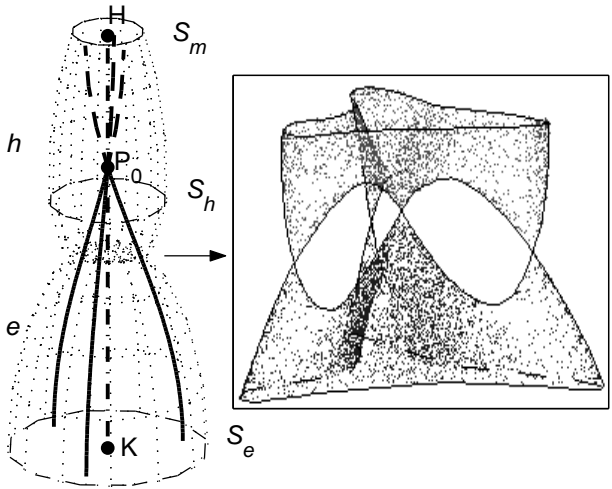}
\caption{\label{fig2} Sketch of the Fermi surface (a half of it)
and of the band contact lines in graphite. The accidental contact
of the bands $\varepsilon_2({\bf k})$ and $\varepsilon_3({\bf k})$
occurs along the solid lines, while the dashed lines mark the
accidental contact of the bands $\varepsilon_3({\bf k})$ and
$\varepsilon_4({\bf k})$. The same bands are in contact along the
HKH axis due to the symmetry of the crystal. All the lines merge
at the point $P_0$. Shown are also the maximum cross sections of
the electron ($S_e$) and hole ($S_h$) majorities and of the hole
minority ($S_m$) for the magnetic field along the HKH axis. A part
of the Fermi surface where the electron and hole majorities touch
is presented in an enlarged scale on the right; the band-contact
lines pass through the conical features of the Fermi surface.
 } \end{figure}   %%%%%%%%%%%%%%%%%%%%%%%%%%%%%%%%%%%%%%%%%%

It is seen from Fig.~1 that two of the four bands are degenerate
along the HKH edge of the Brillouin zone. In the interval from K
to the point $P_0$ defined by the condition $E_1(\xi)= E_3(\xi)$
(i.e., by the equality $\cos\xi \approx |\Delta|/2\gamma_1 \approx
0.01$) these two bands are $\varepsilon_2({\bf k})$ and
$\varepsilon_3({\bf k})$, while from $P_0$ to the point H they are
$\varepsilon_3({\bf k})$ and $\varepsilon_4({\bf k})$. The contact
of these bands is caused by the symmetry of the crystal. The
change of the band degeneracy at $P_0$ can be understood if one
takes into account a small spin-orbit interaction in SWM model
(and then makes the interaction tend to zero). However, we
emphasize here that as it follows from Eqs.~(\ref{5}) - (\ref{7}),
there are three additional contact lines of the same bands which
are also located near the edge HKH, Fig.~2. The contact of the
bands in these lines is accidental, \cite{H} i.e., is not caused
by the symmetry of the crystal. It is due to the so-called
trigonal warping \cite{Br} of the Fermi surface which is
characterized by the parameter $\gamma_3$. In the vicinity of the
point K these lines can be approximately found from the equations:
\begin{eqnarray}\label{8}
 \sin 3\alpha=-1,\ \ \
 \zeta \approx \frac{\sqrt 3}{4}\frac{\gamma_1 \gamma_3}
 {\gamma_0^2}\cos^2\xi .
\end{eqnarray}
Thus, the situation in graphite can be described as follows: The
four contact lines of the bands $\varepsilon_2({\bf k})$ and
$\varepsilon_3({\bf k})$ come to the point $P_0$ from the one side
of the HKH axis, and the four contact lines of the bands
$\varepsilon_3({\bf k})$ and $\varepsilon_4({\bf k})$ come to this
point from the opposite side, and all these lines merge at the
point $P_0$. It is essential that in the vicinity of all these
lines the band splitting is linear in a deviation of ${\bf k}$
from the lines.

In graphite the electrons of the band $\varepsilon_3({\bf k})$
give rise to the majority electron group, while the holes of the
band $\varepsilon_2({\bf k})$ make up the hole majority. All the
contact lines of the bands $\varepsilon_2({\bf k})$ and
$\varepsilon_3({\bf k})$ lie under the Fermi surface of these
majority groups. When the lines pass from the electron part of the
surface to its hole part, the conical features of the Fermi
surface, the so-called ``outrigger'' pieces, \cite{Br} appear,
Fig.~2. The lines are just axes of these four pieces connecting
the electron and hole parts.

As was shown in our paper, \cite{fnt} in the magnetic field $H$
parallel to a band-contact line the splitting of the Landau levels
for the electron states near the line is proportional to $\sqrt H$
rather than to $H$. In the case of graphite these levels are known
as the so-called ``leg levels'' investigated by Dresselhaus.
\cite{Dr} The existence of the leg levels (and, in fact, of the
band-contact lines in graphite) was confirmed by magneto-optical
experiments. \cite{Pl,D} Interestingly, the well-known large
diamagnetism of graphite \cite{Mc1} is also caused by electron
states near the band-contact lines. \cite{MS}

\begin{table}
\caption{\label{TII}Values of the parameters (eV) of the SWM
model. \cite{Br}}
\begin{ruledtabular}
\begin{tabular}{cr|lr}
$\gamma_0$ & $3.16\pm 0.05$ & $\gamma_4$ & $0.044\pm 0.024$ \\
$\gamma_1$ & $0.39\pm 0.01$ & $\gamma_5$ & $0.038\pm 0.005$ \\
$\gamma_2$ & $-0.020\pm 0.002$ & $\Delta$ & $-0.008\pm 0.002$ \\
$\gamma_3$ & $0.315\pm 0.015$ & $\varepsilon_F$ & $-0.024\pm 0.002$ \\
\end{tabular}
\end{ruledtabular}
\end{table}

\section{Discussion}

When the magnetic field $H$ is directed along the $z$ axis, the
maximum electron cross section in $k_z$ is located at $\xi=0$,
while the maximum cross section of the hole majority is between
the points K and $P_0$, viz., at $\cos \xi \approx \pm
(\varepsilon_F/6\gamma_2)^{1/2}\approx 0.45$ where $\varepsilon_F$
is the Fermi energy in graphite, see Fig.~2. Thus, both these
cross sections are penetrated by the four band-contact lines.
However, an {\it even} number of the band-contact lines do not
change \cite{prl,c1} the usual value $\gamma=1/2$. Thus, we find
$\gamma=1/2$ for the maximum cross sections of the majority
groups, which agrees with the experimental results of
Refs.~\onlinecite{S,W}.

We now discuss briefly the value of $\gamma$ for the minority
group. For the parameters presented in Table II, the hole minority
is located near the point H and it results from the band
$\varepsilon_1({\bf k})$. At this point the minority and the hole
majority produced by the band $\varepsilon_2({\bf k})$ have equal
cross sections when the magnetic field is along the HKH axis.
Since no contact lines of the bands $\varepsilon_1({\bf k})$ and
$\varepsilon_2({\bf k})$ penetrate this common cross section, one
might expect to find the usual value $\gamma=1/2$ in this case.
However, the semiclassical approximation which is used in deriving
Eqs.~(\ref{1}) and (\ref{2}) fails for the hole orbits
corresponding to this cross section since for this approximation
to be valid, the orbits must be sufficiently far away from each
other. The analysis carried out beyond the scope of the
semiclassical approximation \cite{W} led to $\gamma=0$ and
$\delta=0$ for the ``degenerate'' orbit. In experiments this orbit
is ascribed to the hole minority, and the phase $\phi_m$ measured
in Ref.~\onlinecite{W} agrees with these $\gamma$ and $\delta$,
see Table I.

In Ref.~\onlinecite{K} a new method was developed to determine the
phase $\phi$ of the de Haas -van Alphen oscillations of the
magnetic susceptibility. The appropriate results for $\phi$ and
$\gamma$ in graphite are presented in Table I. However, authors of
Ref.~\onlinecite{K} implied in their analysis of $\gamma$ that the
sign of $\chi$ in formula (\ref{3}) is positive in the case of
electrons and negative for holes. This is not correct; the sign is
always positive. A re-examination of the derivation of the
Lifshits-Kosevich formula \cite{LK} proves this statement.
\cite{c2} With this in mind we have corrected $\delta$ and
$\gamma$ of Ref.~\onlinecite{K}, and the obtained results are also
presented in Table I.

For the hole minority and for the electron majority \cite{c3}  the
corrected results coincide with those of Williamson {\it et
al}.\cite{W} (but $\delta_m=0$ can be caused by the
above-mentioned degeneracy of the hole orbits rather than by the
two-dimensional spectrum of the hole minority). For the hole
majority the phases $\phi_h$ measured in Refs.~\onlinecite{W} and
\onlinecite{K} disagree. The phase $\phi_h=\pi$ obtained by
Luk'yanchuk and Kopelevich \cite{K} means that either the spectrum
of these carriers is two dimensional, %and hence the values of the
%parameters \cite{Br} of the SWM model require a reconsideration,
or if $\delta=-\pi/4$, one obtains $\gamma=3/8$. However, in the
semiclassical approximation, $\gamma$ can be equal to $1/2$ or to
$0$ only. \cite{prl,zhetf} Intermediate values can occur in
situations close to the magnetic breakdown. \cite{A} In principle,
such the situation is possible for the SWM model, but it does not
occur for the parameters presented in Table II.

The parameters of Table II correspond to three dimensional
spectrum of graphite and lead to a consistent description of the
experimental data \cite{S,T,M,D,Me1,Me2} obtained many yeas ago.
However, Luk'yanchuk and Kopelevich \cite{K} used the highly
oriented pyrolytic graphite (HOPG) with very high ratio of the
out-of-plane to basal-plane resistivities ($\sim 5\times 10^4$),
and in this sample, quantum-Hall-effect features were observed
which indicate a quasi two dimensional nature of this HOPG.
\cite{K1} It was also argued \cite{K2} that in similar samples of
HOPG an incoherent transport occurs in the direction perpendicular
to the graphite layers, and the three dimensional spectrum of
carriers seems to fail. If this conclusion is valid only for the
hole majority, it could explain the above-mentioned disagreement.
This also means that the parameters of SWM model should be
reconsidered to describe the spectrum of such HOPG.

To conclude, the phases of the de Haas - van Alphen oscillations
in graphite were measured in Refs.~\onlinecite{S,W,K}. The data of
Refs.~\onlinecite{S,W} can be completely explained in the
framework of the known band structure of graphite \cite{Br} if one
takes into account that four band-contact lines exist near the HKH
edge of its Brillouin zone. The data of Luk'yanchuk and Kopelevich
\cite{K} obtained for HOPG disagree with the experimental results
of Refs.~\onlinecite{S,W} for one of the two large cross sections
and probably imply that a reconsideration of the energy-band
parameters for such HOPG is required.

{}


\begin{thebibliography}{}
\bibitem{c1} Although each line adds $0.5$ to
$\gamma$, but within the semiclassical approximation $\gamma$ is
defined up to an integer.
\bibitem{c2} Note that for holes, the integration
$\int_0^{\infty}dn$ in Ref.~\onlinecite{LK} is rearranged to the
integration over the energy from $0$ to $-\infty$ rather than from
$0$ to $\infty$.
\bibitem{c3} Note that in Ref.~\onlinecite{K} the frequencies $\nu$
for the electron and hole majorities are interchanged as compared
to popular belief \cite{Br}; see Table I.
\end{thebibliography}
\end{document}